# Modeling and Design of Integrated Iontronic Circuits Based on Ionic Bipolar Junction Transistors

Soichiro Tottori[1,2,*] and Rohit Karnik[1]

[1]*Department of Mechanical Engineering, Massachusetts Institute of Technology, Cambridge, Massachusetts 02139, USA*

[2] *Graduate School of Engineering, Tohoku University, Sendai, Miyagi 980-8579, Japan*

* soichiro.tottori.e2@tohoku.ac.jp

Biological systems rely on ions and molecules as information carriers rather than electrons, motivating the development of devices that interface with biochemical systems for sensing, information processing, and actuation via spatiotemporal control of ions and molecules. Iontronics aims to achieve this vision by constructing devices composed of ion-conducting materials such as polyelectrolyte hydrogels, but advancing beyond simple single-stage circuit configurations that operate under steady-state conditions is a challenge. Here, we propose and model more complex ionic circuits, namely bistable flip-flop and ring oscillators, consisting of multiple ionic bipolar junction transistors (IBJTs). We begin by modeling and characterizing single IBJTs using both a simplified one-dimensional Nernst-Planck model and a more-detailed two-dimensional Poisson-Nernst-Planck model, elucidating the effects of geometry, size, and fixed charge on the IBJT performance and response time. The one- and two-dimensional models exhibit good agreement, indicating negligible transverse inhomogeneities. Additionally, these models show that reducing the base width improves current amplification, a behavior analogous to electronic BJTs. Building on this understanding, by using the IBJTs as voltage inverters and buffers, we design and model more complex ionic circuits that dynamically change their states in response to ionic signals. Specifically, we demonstrate that the ionic flip-flop retains one-bit memory and that the ring oscillator achieves autonomous periodic self-oscillation without an external clock. Our work provides a foundation for designing dynamic iontronic circuitry using ionic conductors, enabling biochemical signal processing and logic operations based on ionic transport.

## I. INTRODUCTION

Biological systems utilize various ions and molecules as information carriers rather than electrons. Technologies to directly sense or manipulate such biochemical entities in aqueous environments are therefore important for interfacing with biological systems, including for information exchange, sensing, and actuation. Various devices have been developed so far for this purpose, including organic electrochemical transistors (OECTs) and electrophoretic ion pumps [1-5]. These devices have potential applications including sensing of various biological signals, stimulating cells or tissues, delivery of therapeutics, and construction of engineered living systems [6,7]. In recent years, the field concerned with the understanding and design of such devices has been collectively referred to as iontronics [8,9].

Controlling the transport of ions and charged molecules in confined environments, such as nanochannels or nanopores [10-12] and porous materials [13-15], is critical in iontronics. Drawing analogy from the conventional electronics, various iontronic components have been developed, including diodes [16-20], field-effect transistors [21,22], bipolar junction transistors [23-26], ion-to-ion amplifiers [27], and power sources [28-30]. Simple logic gates, such as AND, OR, NOR, NAND, and NOT, have also been implemented by combining multiple components [31-35]. However, the integration of iontronic components to design iontronic circuits has mainly focused on relatively simple single-stage configurations, with device modeling performed largely under steady-state conditions. Extending these systems to circuits comprising multiple components will enable the realization of dynamic and more complex behaviors.

Recently, a one-dimensional modular framework for modeling ionic circuits was proposed, providing an important step toward systematic circuit-level design [36].

Here, we propose and numerically model dynamic, fully iontronic systems composed of multiple ionic bipolar junction transistors (IBJTs) connected in series. We begin by characterizing individual IBJTs consisting of cation- and anion-selective regions using finite element simulations based on the full two-dimensional Poisson–Nernst–Planck (PNP) equations. This allows us to systematically explore geometric parameters and identify key design principles for effective ion amplification. By adjusting the position of the base region, we construct functional circuit elements such as voltage inverters and buffers. Extending the model to time-dependent regimes, we integrate these elements into dynamic iontronic architectures, including bistable flip-flops and ring oscillators. Notably, the proposed ring oscillator operates without explicit capacitive components, which are challenging to practically implement in iontronic systems. Furthermore, we introduce a buffering method that enables ion species conversion and removes the need to maintain constant ion concentration conditions between adjacent inverters, thereby enabling sustained operation. Our work provides a framework for the design and implementation of entirely ion-based information processing systems.

## II. METHODS
### A. Two-dimensional (2D) Poisson-Nernst-Planck (PNP) model

We begin by considering an IBJT, as schematically illustrated in Fig. 1(a). Analogous to electric BJTs, an IBJT consists of a collector, base, and emitter, but it is made of ionically conducting porous materials, such as hydrogels or nanochannel networks. In Fig. 1(a), the collector and emitter are assumed to possess fixed positive charges, while the base carries fixed negative charges. In the two-dimensional (2D) Poisson-Nernst-Planck (PNP) model for ion transport, the device is modeled as consisting of 2D regions with fixed volumetric charge densities, as shown in Fig. 1(b). For simplicity, we assume that the fixed charges in the cation- and anion-selective regions are uniformly distributed. The electric potential $\phi$ and fixed volume charge density $\rho$ are related by the Poisson equation

$$\epsilon \nabla^2 \phi = -F \sum_i z_i c_i - \rho, \tag{1}$$

where $\epsilon$, $F$, $z_i$, and $c_i$ are the permittivity of the medium, Faraday constant, valence, and concentration of each ion, respectively. The typical orders of magnitude for $c_i$ and $\rho$, based on previous experimental studies on IBJTs, are approximately $c_i \sim 1 \times 10^2$ mM and $|\rho| \sim 1 \times 10^8$ C/m$^3$ [24,33]. The ionic flux $\mathbf{J}_i$, driven by both diffusion and electromigration, is given by

$$\mathbf{J}_i = -D_i \nabla c_i - z_i \mu_i F c_i \nabla \phi, \tag{2}$$

where $D_i$ and $\mu_i$ are the diffusion coefficient and mobility of each ion, respectively. The mobility and diffusion coefficient are related by the Nernst-Einstein relation

$$\mu_i = \frac{D_i}{RT}, \tag{3}$$

where $R$ and $T$ are the universal gas constant and temperature, respectively. We use the diffusion coefficients of potassium chloride, $D_1 = 1.957 \times 10^{-9}$ m$^2$/s for K$^+$, and $D_2 = 2.032 \times 10^{-9}$ m$^2$/s for Cl$^-$, throughout all simulations. Note that here we neglect the advection term (primarily caused by electroosmotic flow) for simplicity. The conservation of each ion species is described by

$$\frac{dc_i}{dt} = -\nabla \cdot \mathbf{J}_i, \tag{4}$$

where $t$ is time.

The boundary conditions at the interfaces between two charged regions are the continuity of ion concentration, electric potential, and flux. At the interfaces between the charged regions and external terminals, we impose constant ion concentrations and electric potentials. These coupled equations are solved using a commercial finite element simulation package (COMSOL Multiphysics, ver. 5.6) for the 2D geometries shown in Fig. 1(b). The ionic current is calculated by

$$I = F \sum_i z_i \int \mathbf{J}_i \cdot \mathbf{n}\, dl, \tag{5}$$

where the integral is taken over the cross-sectional length, and **n** is the unit normal vector perpendicular to the integration path.

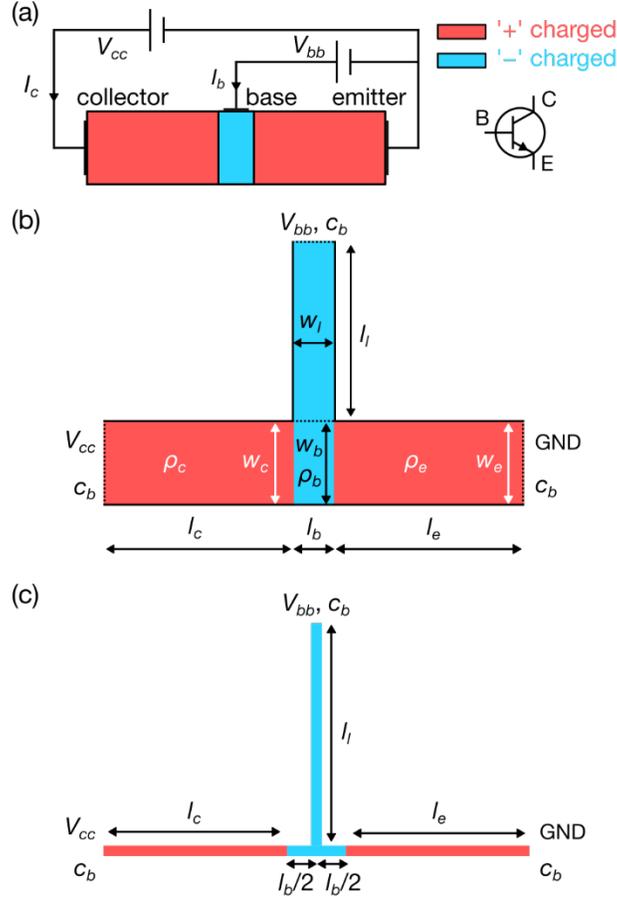

Fig. 1. Ionic bipolar junction transistor (IBJT). (a) Schematic illustration of an *npn*-type IBJT, where positively and negatively charged regions are shown in red and blue, respectively. (b) Schematic illustration of the 2D simulation domain and boundary conditions. (c) Schematic illustration of the 1D simulation domain and boundary conditions.

### B.  One-dimensional (1D) Nernst-Planck (NP) model

When the ion concentration and electric potential profiles are approximately uniform along the transverse (width) direction of the charged regions of the device, the model can be simplified to a 1D system, as shown in Fig. 1(c), where each 2D charged region is replaced by a 1D charged line segment. To enforce conservation of each ionic species at junctions where multiple segments meet, we impose the condition

$$\sum_k w_k J_{i,k} = 0, \tag{6}$$

where the summation runs over all segments $k$ connected to a given junction. Here, $w_k$ and $J_{i,k}$ denote the width and the ionic flux density of segment $k$, respectively.

To simplify the model, we apply the Donnan potential at the interfaces between regions with different fixed charge densities, assuming local Donnan equilibrium. Although the applied voltage across the system is relatively high, the space charge region is negligibly thin compared to the minimum feature size of the system, justifying this approximation. The validity of this approximation will be discussed in the Results section by comparing with the 2D PNP solutions. In this study, the base length $l_b$, which corresponds to the smallest geometric feature of the system, must be sufficiently large compared to the length scale of the space charge region at the junction. Under these conditions, electroneutrality can be assumed everywhere as

$$\sum_i z_i F c_i + \rho = 0. \tag{7}$$

In dimensionless form, this becomes

$$\sum_i z_i c_i^* + \rho^* = 0, \tag{8}$$

where $c_i^* = c_i/c_b$ is the dimensionless electrolyte concentration, $\rho^* = \rho/(c_b F)$ is the dimensionless volume charge density, and $c_b$ is the bulk ion concentration. For simplicity, we consider only monovalent ions in this work. The dimensionless volumetric charge density $\rho^*$ corresponds to the Dukhin number $|\rho^*| = \mathrm{Du}$, which characterizes the relative conductivity contribution of fixed charges to the bulk conductivity. At the junctions of different fixed charge regions, the electric potential shifts by the Donnan potential, given by

$$V_D = \frac{RT}{Fz_i} \ln\left(\frac{c_b}{c_i}\right). \tag{9}$$

In dimensionless form, this becomes

$$V_D^* = \frac{1}{z_i} \ln\left(\frac{1}{c_i^*}\right), \tag{10}$$

where $V_D^*$ is the Donnan potential nondimensionalized by the thermal voltage, $\phi_{\mathrm{th}} = RT/F$ (~26 mV). The ionic flux density is nondimensionalized by $D_0 c_b/L$, resulting in

$$J_i^* = -D_i^* \nabla^* c_i^* - z_i D_i^* c_i^* \nabla^* \phi^*, \tag{11}$$

where $D_0$ and $L$ denote the characteristic diffusion coefficient and length scale, respectively, and $D_i^* = D_i/D_0$ are the dimensionless diffusion coefficients. The dimensionless Nernst-Planck equation is thus

$$\frac{dc_i^*}{dt^*} = -\nabla^* \cdot J_i^*, \tag{12}$$

where $t^*$, $\phi^*$, and $\nabla^*$ are defined as $t^* = D_0 t/L^2$, $\phi^* = \phi/\phi_{\mathrm{th}}$, and $\nabla^* = L\nabla$, respectively. The dimensionless ionic current is

$$I^* = w^* \sum_i z_i J_i^*. \tag{13}$$

The boundary condition imposed at the interface between the charged segments and the external terminals are given by continuity conditions for concentrations and electrochemical potentials, expressed as

$$c_{1,a}^* c_{2,a}^* = c_{1,b}^* c_{2,b}^*, \tag{14}$$

and

$$\phi_a^* - V_{D,a}^* = \phi_b^* - V_{D,b}^*, \tag{15}$$

where the subscripts $a$ and $b$ denote each side of the interface between different charged segments. The symbols are summarized in Table II in Appendix.

## III. RESULTS
### A. Operating principles of ionic bipolar junction transistors (IBJTs)

An IBJT operates on a principle analogous to that of a traditional bipolar junction transistor (BJT), with the key distinction that it relies on ionic conduction rather than electronic conduction. In the IBJT structure shown in Fig. 1(a), the collector and emitter are positively charged, while the base is negatively charged. When a voltage is applied between the base and the emitter, positive ions migrate from the base toward the base-emitter junction, while negative ions move from the emitter toward the same junction. This accumulation of ions enhances the ionic concentration at the collector-base junction, effectively lowering the interfacial resistance and facilitating ion transport across the junction. As a result, a large flow of negative ions is induced from the emitter to the collector, enabling signal amplification via ionic transport. Through this mechanism, the IBJT can function as an amplifier of ionic signals, making it a fundamental building block for ionic circuitry. Key performance metrics for evaluating IBJT operation include the current gain, leak current (off-state current), and response time.

In contrast to previous studies [24-26,33], our design employs a negatively charged base instead of a neutral junction between the collector and emitter to suppress off-state ionic leakage current. The 2D simulation domain and the boundary conditions are schematically shown in Fig. 1(b). In this study, we define the 'collector/base/emitter lengths' ($l_{c,b,e}$) as the horizontal lengths of the corresponding regions and the 'collector/base/emitter widths' ($w_{c,b,e}$) as the vertical lengths with the planar IBJT geometry, which differ from those used in conventional BJT structures. This terminology is employed throughout the paper for clarity. Assuming that the ion concentration and potential do not change considerably across the width of the charged regions, we can also simplify the 2D model to a 1D representation, as shown in Fig. 1(c).

The IBJT's on/off response was characterized by applying a constant voltage at the collector ($V_{cc} = 0.5$ V or $V_{cc}^* \approx 19.5$) and variable voltages at the base ($V_{bb} = 0 \sim 0.5$ V or $V_{bb}^* = 0 \sim 19.5$). We set the ion concentration $c_i = 1 \times 10^2$ mM and the fixed charge density $|\rho| = 1 \times 10^8$ C/m$^3$, which are the typical values from previous experimental systems [24-26,33]. The resulting Dukhin number is thus Du $\approx$ 10. Other parameters used in the 2D and 1D models are summarized in Table 1 in Appendix.

In Fig. 2(a)–2(c), the nondimensionalized electric potential and ion concentration profiles from the collector end to the emitter end are plotted as a function of dimensionless position $x^*(=x/l_c)$, where, $l_c$ is the length of the collector. The 1D NP models show discontinuous jumps in potential at each interface due to the Donnan potentials, while the 2D PNP models show continuous but steep potential gradients, where electroneutrality is not conserved in the space charge region. As long as the base length $l_b$ is large enough in comparison to these spatially charged regions, the 1D model shows good agreement with the 2D PNP simulation results. Similar to conventional electronic BJTs, when $V_{bb}^* = 0$, most of the potential drop occurs at the collector-base interface, which act as a reversed-biased diode. As the base voltage is increased, ion concentrations in the base and the emitter increase, as the base-emitter interface becomes a forward-biased diode. This local ion accumulation reduces the potential drop at the collector-base interface, as shown in Fig. 2(b) and 2(c). In Fig. 2(d), 2D maps of dimensionless equivalent ion concentrations $\sqrt{c_1 c_2}/c_b$ are plotted for the region around the base. This equivalent concentration corresponds to the hypothetical salt concentration in an uncharged region in local Donnan equilibrium with the point under consideration, and is found to be highly homogeneous along the width direction, which explains the strong agreement between the 1D NP and 2D PNP models. In Fig. 3, the ionic currents through the collector and base are plotted as a function of the applied voltage at the base. Again, the 1D NP and 2D PNP models show good agreement, both capturing the ionic current amplification behavior of the IBJT.

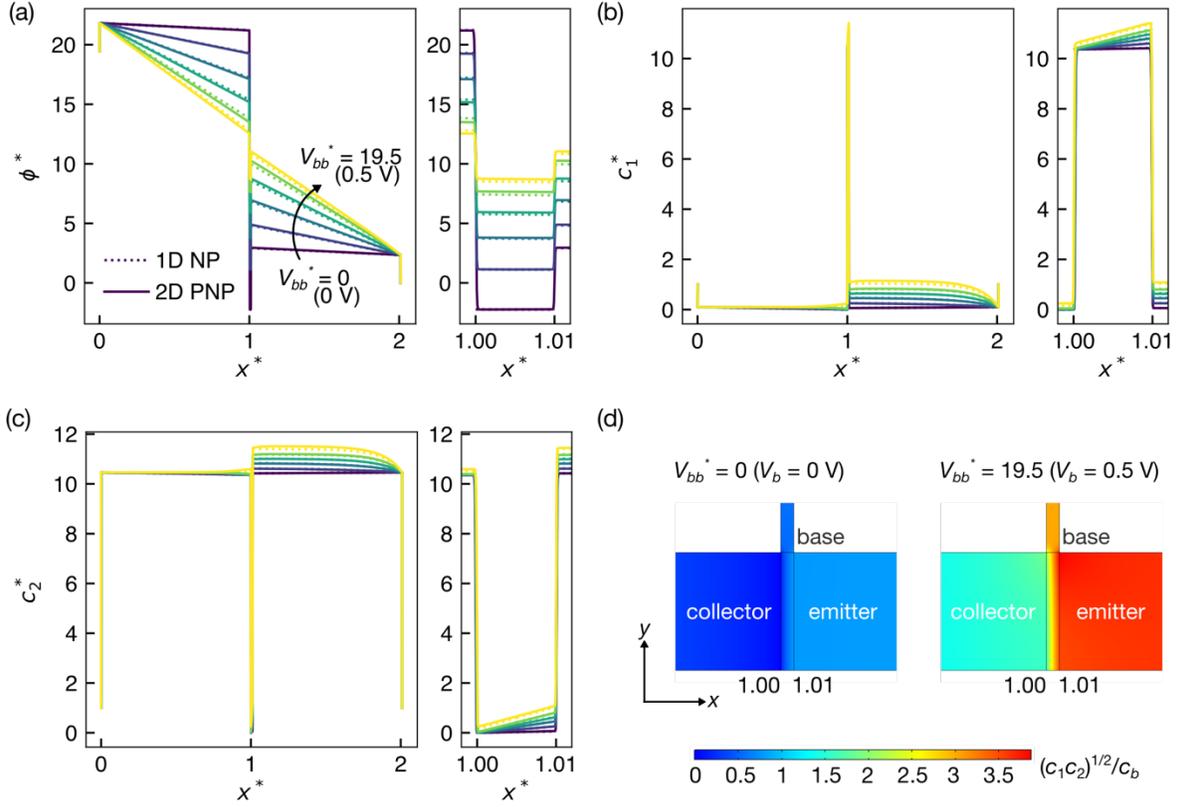

Fig. 2. Characterization of a single IBJT. (a–c) Potential and concentration profiles from the collector to the emitter for various base voltages from 0 to 0.5 V in increments of 0.1 V. Insets show zoomed-in views around the junctions. The supplied voltage is 0.5 V. (a) Potential profiles. (b) Cation concentration profiles. (c) Anion concentration profiles. (d) Color maps of dimensionless bulk equivalent concentration $\sqrt{c_1 c_2}/c_b$ at the base. Parameters used are summarized in Table I in Appendix.

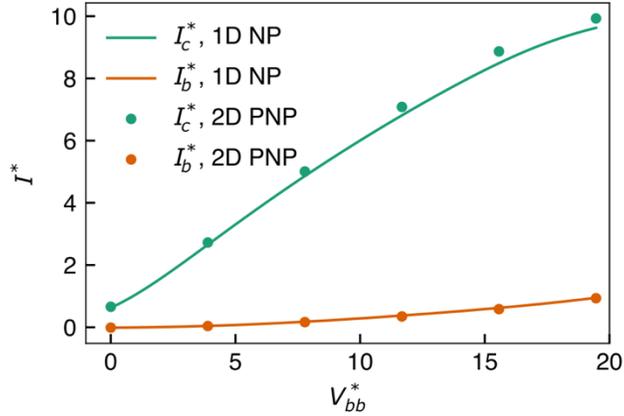

Fig. 3. Collector current $I_c^*$ and base current $I_b^*$ as a function of the base voltage $V_{bb}^*$. The supply voltage $V_{cc}^*$ is 19.5. Parameters used are summarized in Table I in Appendix.

## B. Parameter study of IBJTs

Having established a basic understanding of how IBJTs operate, we now perform a parametric study to systematically tune their operational characteristics. The key performance metrics are: (1) the current amplification ratio, (2) the on/off ratio, and (3) the response time. Specifically, we define: (1) the current amplification ratio as $\frac{dI_c^*}{dI_b^*}|_{V_{bb}^*=0}$, (2) the on/off ratio as $\frac{I_c^*|_{V_{bb}^*=V_{cc}^*}}{I_c^*|_{V_{bb}^*=0}}$, and (3) the rise time $\tau_r^*$ and fall time $\tau_f^*$ as the time intervals during which the collector current increases from 0% to 90%, and decreases from 100% to 10% of its saturated value $I_{c,(t\to\infty)}^*$, respectively. The off-state current is defined as $I_c^*|_{V_{bb}^*=0}$, and $I_c^*|_{I_b^*=0}$ corresponds to the floating base condition. In Fig. 4(a), the collector current $I_c^*$ is plotted as a function of the base current $I_b^*$ for various base lengths $l_b^*$. The applied collector voltage $V_{cc}$ is fixed at 0.5 V ($V_{cc}^* \approx 19.5$), while base voltage $V_{bb}$ is swept from 0 to 0.5 V ($V_{bb}^* = 0\sim19.5$). The solid lines represent the results of the 1D PN models, and the dots show the results of the 2D PNP models using the geometrical parameters in Table I in Appendix. The base lead length $l_l^*$ is set to 1 for all cases, except for $l_b^* = 0.01$, where $l_l^*$ takes values 0.5, 1, 2 and these cases are plotted with light-colored dots. The data show that the $I_c^*$–$I_b^*$ curves are largely independent of the base lead lengths. Although the base lead geometry affects the relationship between the base voltage $V_{bb}^*$ and the base current $I_b^*$, the collector current is primarily governed by the base current. When the base lead length $l_l^*$ is sufficiently long, the base lead resistance and applied base voltage determine the base current, which in turn sets the magnitude of the collector current. Therefore, it is possible to tune the working voltage range of the IBJT without changing the current amplification ratio. The current amplification ratio can be increased by reducing base length $l_b^*$, due to the resulting steeper ion concentration gradient in the base. However, shorter base length also deteriorates the on/off ratios, as the reverse-biased diode effect at the collector junction becomes weaker.

Figure 4(b) shows the effect of the applied collector voltage $V_{cc}^*$ on the $I_c^*$–$I_b^*$ characteristics. The on/off ratio improves with higher $V_{cc}^*$, since the off-state current remains nearly constant while the on-state current increases approximately linearly with $V_{cc}^*$. The current amplification ratio remains essentially unchanged, although a higher base current is required to reach saturation at larger $V_{cc}^*$.

Figure 5(a) shows the time response of the collector and base currents upon application of a square wave pulse to the base input. The response time scales as $\sim L^2$, as expected from the dimensionless time $t^* = D_0 t/L^2$, which applies to both the rise and fall phases, as shown in Fig. 5(b). This behavior is consistent with previous work on voltage-driven iontronic systems [37], indicating that such diffusion-like scaling is robust across different circuit configurations.

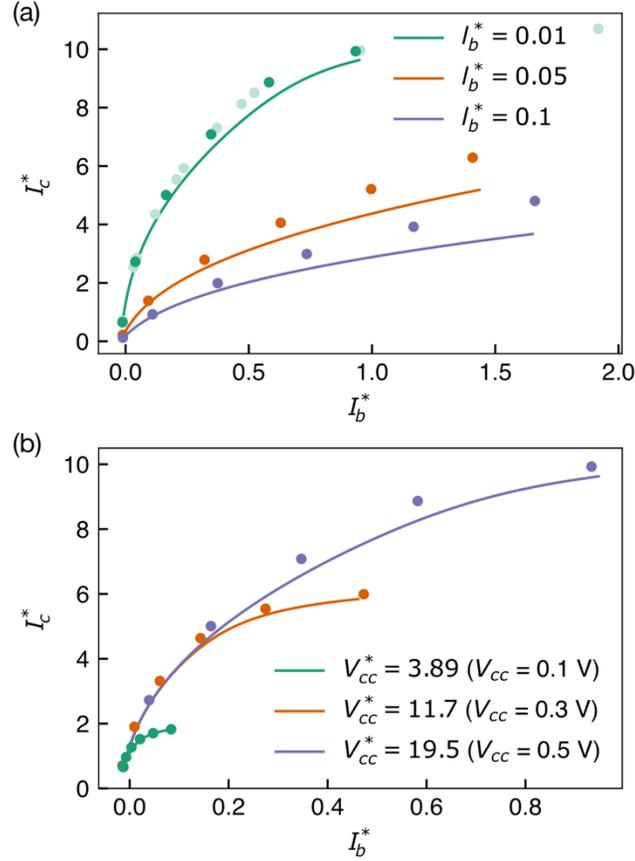

Fig. 4. IBJT performance. (a) Collector currents $I_c^*$ as a function of base currents $I_b^*$ for various base lengths $l_b^*$. Dots represent the results of the 2D PNP simulation, and solid lines indicate the 1D NP model results. The supply voltage is set to $V_{cc} = 0.5$ V, or equivalently $V_{cc}^* \approx 19.5$. The base voltages are swept from 0 to $V_{cc}^*$. The base lead length $l_l^*$ is set to 1 for all cases, except for $l_b^* = 0.01$, where $l_l^*$ takes the values 0.5, 1, and 2 (the cases with $l_l^* = 0.5$ and 2 are shown as light-colored dots). (b) Collector currents $I_c^*$ as a function of base currents $I_b^*$ for various applied voltages $V_{cc}^*$. The base length $l_b^*$ is set 0.01. Dots represent the 2D PNP simulation results, and solid lines indicate the 1D NP model results. The base voltage is varied from 0 to $V_{cc}^*$. Parameters used are summarized in Table I in Appendix.

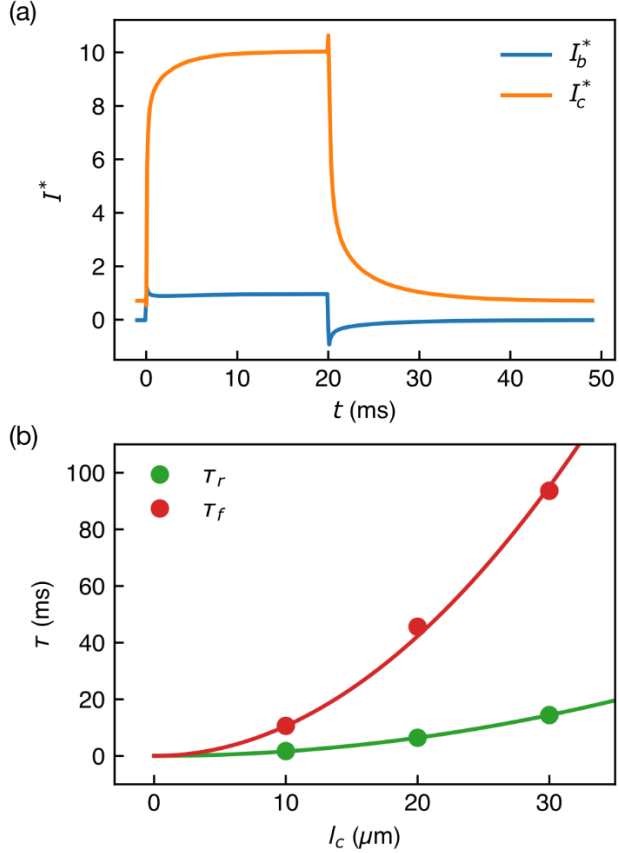

Fig. 5. Time response of the IBJT. (a) Collector and base currents as a function of time. The geometry is identical to that used in Fig. 2 and Fig. 3. The supply voltage $V_{cc}^*$ is to 19.5, and the base voltage $V_{bb}^*$ is held at 19.5 during $0 \leq t \leq 20$ ms, and 0 otherwise. (b) Rise and fall times ($\tau_r$ and $\tau_f$) as a function of characteristic length $l_c$. Solid lines represent fits to $\tau \propto l_c^2$. The geometry with $l_c = 10$ μm corresponds to the case shown in (a). Rise and fall times are defined as the durations for the current to increase from 0% to 90% and to decrease from 100% to 10%, respectively. Parameters used are summarized in Table I in Appendix.

### C. Voltage inverter and buffer

Building on the characterization of symmetric IBJTs, voltage inverters and buffers were implemented by tuning the resistance ratio between the collector and emitter, analogous to conventional electronic circuits, as illustrated in Fig. 6(a). The configurations correspond to: (i) an *npn*-inverter, (ii) an *npn*-buffer, (iii) a *pnp*-buffer, and (iv) a *pnp*-inverter, respectively. The typical potential profiles of the *npn*-inverter with the parameters in Table I are shown in Fig. 6(b). When $V_{in}$ is low, the voltage drop primarily occurs at the collector-base junction. As $V_{in}$ increases, this voltage drop diminishes, and the potential instead decreases gradually along the length of the collector lead. Figure 6(c) illustrates the potential profiles of the *npn*-buffer. When $V_{in}$ is low, the potential again drops primarily at collector-base junction, keeping $V_{out}$ low. When $V_{in}$ is high, the voltage drop at the collector-base junction vanishes, and $V_{out}$ rises accordingly. In summary, an inverter reverses the input signal: when $V_{in}$ is high $V_{out}$ is low, and vice versa. In contrast, a buffer transmits the input signal unchanged: a high $V_{in}$ produces a high $V_{out}$, and a low $V_{in}$ produces a low $V_{out}$. In the case of iontronics, switching of the charge carrier is an important buffering function that is achieved in these buffering devices. Due to device symmetry, reversing the fixed charges and applied potentials yields equivalent behaviors. In the *npn* configurations (i, ii), the primary voltage drop occurs at the collector junction, while in the *pnp* configurations (iii, iv), it occurs at the emitter junction.

Figure 6(d) shows the 2D PNP simulation results of the output voltage $V_{out}$ as a function of the input voltage $V_{in}$ for four configurations shown Fig. 6(a). The simulation parameters are listed in Table I in Appendix. In conventional electronic BJT inverters, the $V_{out}-V_{in}$ characteristics typically exhibit three distinct regions: cutoff, where $V_{out}$ remains high for small $V_{in}$; linear (or active), where $V_{out}$ decreases approximately linearly with $V_{in}$; and saturation, where $V_{out}$ reaches a low plateau [38]. In contrast, the IBJT inverters presented here lack a cutoff region: $V_{out}$ begins decreasing linearly with $V_{in}$ from the onset. This absence of a cutoff regime is a known challenge in IBJT design, and previous studies have attempted to address it through circuit-level modifications. For example, Tybrandt *et al.* introduced a pull-down voltage terminal to offset the input, thereby artificially imposing a cutoff region in the $V_{out}-V_{in}$ response. This adjustment compensated for the relatively large off-stage leakage currents in their devices due to weak ionic current rectification at the neutral junctions [33]. While this strategy effectively created a cutoff region, this approach is not suitable for sequentially connected IBJT circuits, as ionic currents would leak through the pull-down voltage terminal. To avoid this added complexity, we use the inverter designs in Fig. 6 without further modification.

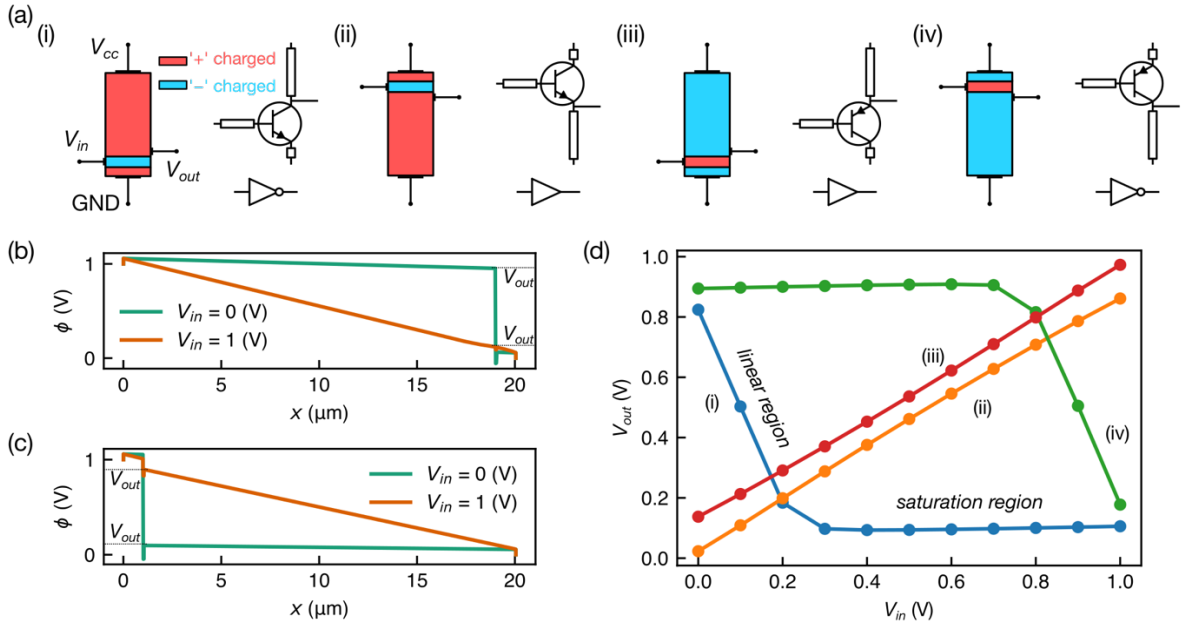

Fig. 6. IBJT-based voltage inverters and buffers. (a) Schematic illustrations of *npn*- and *pnp*-type inverters and buffers: (i) *npn*-type inverter, (ii) *npn*-type buffer, (iii) *pnp*-type buffer, and (iv) *pnp*-type inverter. (b) Potential profiles of the *npn*-type inverter along the collector, base, and emitter. (c) Potential profiles of the *npn*-type buffer along the collector, base, and emitter. (d) Output voltages as a function of input voltage for each configuration. The supply voltage is set to $V_{cc} = 1$ V.

### D. Flip-flop

Connecting an even number of inverters results in a flip-flop circuit. A flip-flop is a type of bistable multivibrator, characterized by two stable states, and can therefore store binary information [38]. We construct an ionic flip-flop using two *npn*-type IBJT inverters, as illustrated in Fig. 7(a). The output terminal extends from the adjacent region of the base, with the same length as the base lead. The total length of the base lead and the output lead length is matched to the base lead length used in the configuration shown in Fig. 6. To avoid ionic current rectification, the interface between the output (input) terminal of the first inverter and the input (output) terminal of the second inverter must maintain a constant ion concentration boundary condition ($c = c_b$), which is assumed in our model. In practice, this condition can be achieved using reservoirs with sufficiently large volumes of electrolyte to maintain a constant ion concentration.

When constant voltages are applied to the two inverters, the system evolves into a bistable state: one output switches to the ON state and the other to the OFF state. Introducing a small asymmetry in the initial conditions, such as a slight time delay in activation, can determine which output turns on. Figure 7(b) shows 2D PNP simulation results of the applied voltages at the two collectors, $V_{cc(1,2)}$, and the corresponding output voltages, $V_{\text{out}(1,2)}$, as functions of time. When both $V_{cc(1,2)}$ are active, the circuit remains bistable. The two states can be switched by briefly turning off and then reactivating the inverter that was initially on.

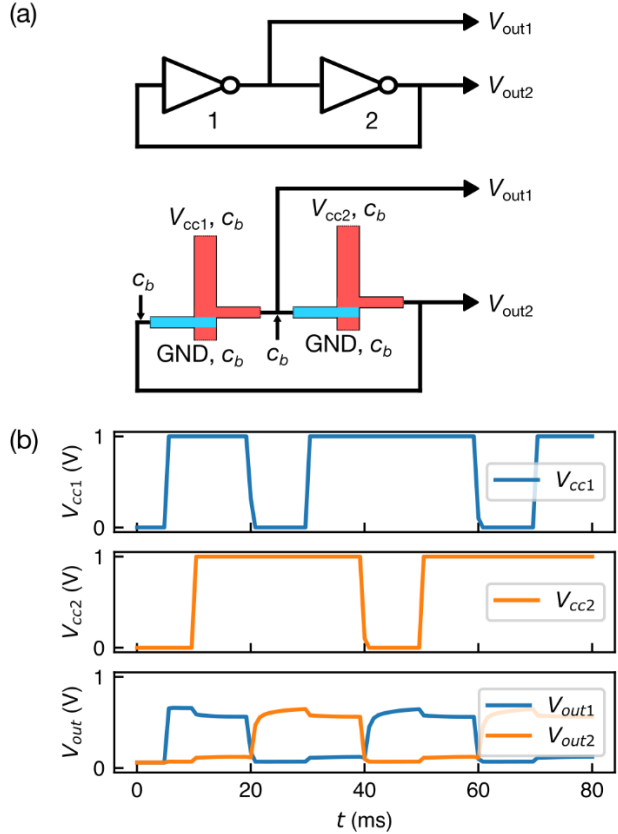

Fig. 7. Ionic flip-flop. (a) Circuit diagram and schematic illustration of an ionic flip-flop based on two *npn*-type inverters. (b) Supplied voltages $V_{cc1,2}$ and output voltages $V_{\text{out}1,2}$ as a function of time. The output state switches when one of the collector terminal voltages is turned off and then on.

### E. Ring oscillator

Connecting an odd number (three or more) of inverters in a loop generally results in a ring oscillator. A ring oscillator is a type of oscillation circuit that generates a periodic signal without requiring an external clock. We again use the *npn*-IBJT inverters to construct an ionic ring oscillator, as illustrated in Fig. 8(a). The electrolyte concentration at the interfaces between the adjacent IBJTs is again set to the bulk concentration $c_b$. To start the oscillation, a supply voltage of $V_{cc} = 1$V is applied simultaneously to each inverter at time $t = 0$, except for the first one, which is given a slight time delay (~ 0.01 ms) to introduce asymmetry. The supply voltages are then held constant at 1 V.

Figure 8(b) shows the output voltages as a function of time for various numbers of stages. We found that in this particular design, the number of stages needs to be at least $n = 5$ for undamped oscillation. The attenuation of oscillation at $n = 3$ arises because the time delay for a signal to propagate through the ring of inverters is insufficient compared to the response time of the individual inverters. By further increasing the number of stages to $n = 7$ and $9$, the propagation delay increases, causing the output voltage to approach its peak and the oscillation frequency to decrease. Figure 8(c) presents the simulation results of the 7-stage ring oscillator at various applied voltages ($V_{cc} = 0.5, 1, 1.5$ V). Both the amplitude and frequency increase with $V_{cc}$, consistent with the behavior observed in electronic ring oscillators.

These results provide a basis for estimating the performance of experimentally realizable devices. Since the response time scales as $\propto L^2$, as shown in Fig. 5(b), for a device with a total length of 2 mm (collector: 1.9 mm, base: 5 μm, emitter: 50 μm), which corresponds to 100 times larger in size, the oscillation time would be approximately $t \sim 10$ s. Therefore, the systems fabricated via methods such as 3D printing [15], photolithography [33], or paper-based techniques [30] could feasibly achieve sub-minute oscillations.

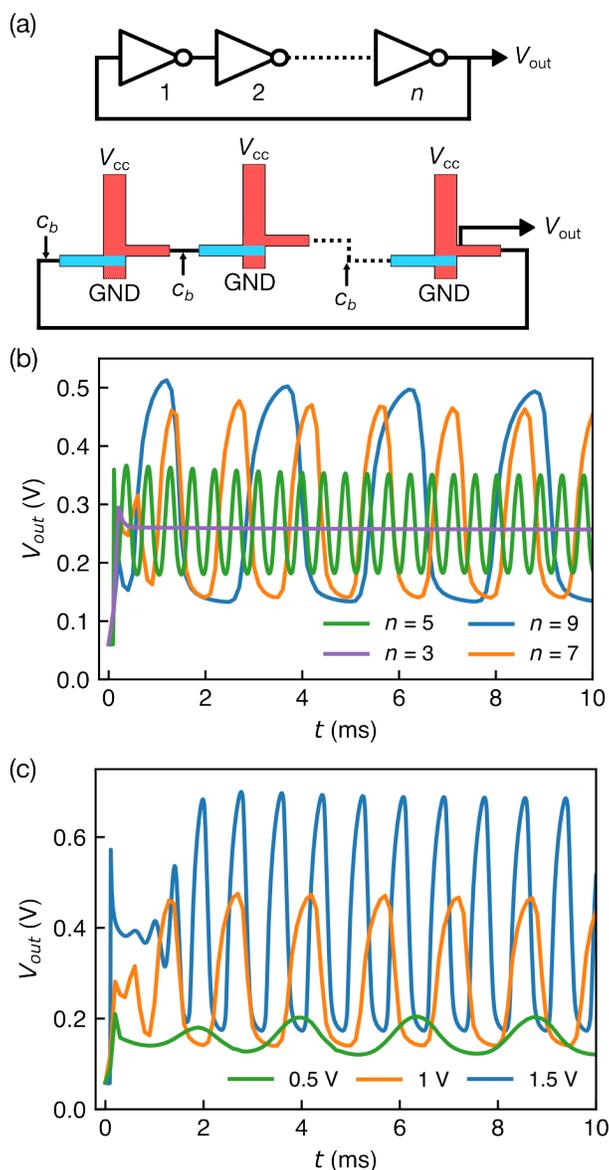

Fig. 8. Ionic ring oscillator. (a) Circuit diagram and schematic illustration of an ionic ring oscillator based on an odd number of *npn*-type inverters. (b) Output voltages as a function of time for various number of stages. The supplied voltage is $V_{cc} = 1$ V. (c) Output voltages as a function of time for various supplied voltages, $V_{cc}$. The number of stages is $n = 7$.

### F.     Ionic circuits without constant ionic concentration reservoirs

In general, ionic circuits, including the flip-flop and ring oscillator described above, require constant concentration boundary conditions at the connections of oppositely charged terminals to avoid the rectification of ionic currents. In practice, such constant concentration conditions can be realized by incorporating sufficiently large reservoirs [33,39], e.g., a solution-impregnated neutral porous material such as hydrogel. However, a potential drawback of relying on constant concentration reservoirs is that the circuit behavior may change over time as the ion concentrations in the reservoirs gradually shift during extended operation. To address this issue, we propose an alternative configuration that eliminates the need for constant concentration reservoirs between adjacent inverters. Specifically, we insert *pnp*-type buffers between *npn*-type inverters for both the flip-flop and ring oscillator, as schematically illustrated in Fig. 9(a). The *pnp*-type buffer converts the anion-dominant output voltage signal from the *npn*-type inverter into a cation-dominant output without altering its potential.

Figure 9(b) shows that the modified flip-flop, incorporating buffers, successfully maintains bistability using two sets of inverters and buffers. When constant voltages are applied to these sets, the system evolves into a bistable state: one output switches to the ON state and the other to the OFF state. The two states can be switched by briefly turning off and on one of the supplied voltage sets. Similarly, in Fig. 9(c), the ring oscillator with buffers incorporated between inverters was also able to oscillate. As shown in the plot in Fig. 9(c), in this configuration, a minimum of 7 stages is required to achieve undamped oscillations. While this buffer-insertion approach adds considerable structural complexity, it demonstrates a viable strategy for expanding design flexibility by inverting the dominant charge carrier, potentially enabling more robust and adaptable ionic circuits.

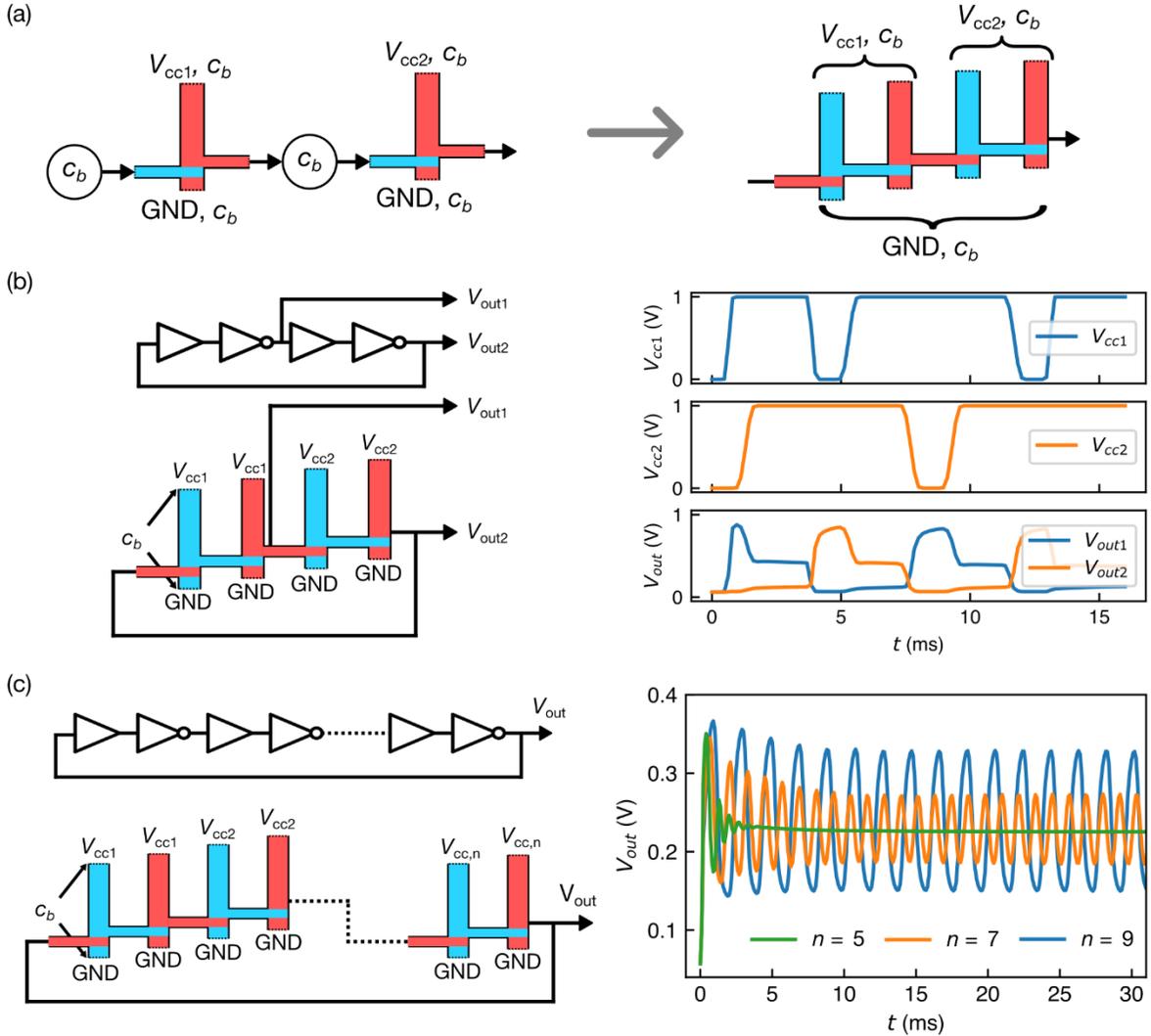

Fig. 9. Ionic circuits with buffers. (a) Schematic illustrating the replacement of constant concentration reservoirs with *pnp*-type buffers inserted between *npn*-type inverters to prevent rectification at oppositely charged interfaces. (b) Ionic flip-flop with buffers: circuit diagram, schematic, and output voltages as a function of time. The output state changes upon sequential switching of the collector terminal voltages. (c) Ionic ring oscillator with buffers: circuit diagram, schematic, and output voltage as a function of time. Undamped oscillation requires a minimum of 7 stages. The supplied voltage is $V_{cc} = 1$ V.

## IV. CONCLUSIONS

In this work, we demonstrated the feasibility of constructing relatively complex ionic circuitry, such as flip-flops and ring oscillators, using 1D and 2D numerical simulations. First, we performed 1D NP and 2D PNP simulations of individual symmetric IBJTs, confirming good quantitative agreement between them. We then systematically characterized the IBJTs in terms of geometric parameters and applied voltages. Second, by shifting the base position, we created *npn*- and *pnp*-type voltage inverters and buffers. Using these components, we implemented flip-flops and ring oscillators with even- and odd-number inverters, respectively. Finally, we showed that inserting *pnp*-type buffers between *npn*-type inverters can eliminate the need for constant concentration reservoirs otherwise required at adjacent connections. Overall, this work establishes that the output signals of IBJTs can be used as the inputs for other IBJTs, paving the way for the design of various more complex ionic circuits. We believe this study offers a practical framework for designing IBJT-based ionic circuitry that could be practically implemented by patterning or additive manufacturing of porous materials such as hydrogels.

## APPENDIX

This appendix provides supplementary details on the simulation parameters and notation used throughout the paper. Table I lists the geometrical parameters used in the simulations. For the 1D and 2D simulations in Figs. 2–5, both dimensional and dimensionless parameters are provided for comparison, while only dimensional parameters are presented for Figs. 6–9. Table II summarizes the symbols used in our models.

TABLE I. Geometrical parameters for the 2D PNP and the 1D NP models.

| Symbol | Description | IBJT in Fig. 2, 3, 5 | IBJT in Fig. 4 | Inverter in Fig. 6, 7, 8 | Inverter in Fig. 9 |
|---|---|---|---|---|---|
| $l_c$ ($l_c^*$) | Collector length | 10 μm (1) | 10 μm (1) | 19 μm | 19 μm |
| $l_b$ ($l_b^*$) | Base length | 0.1 μm (0.01) | 0.1, 0.5, 1 μm (0.01, 0.05, 0.1) | 0.05 μm | 0.05 μm |
| $l_e$ ($l_e^*$) | Emitter length | 10 μm (1) | 10 μm (1) | 1 μm | 0.5 μm |
| $l_l$ ($l_l^*$) | Base lead length | 10 μm (1) | 5, 10, 20 μm (0.5, 1, 2) | 20 μm | 20 μm |
| $w_{c,b,e}$ ($w_{c,b,e}^*$) | Collector/Base/Emitter width | 1 μm (0.1) | 1 μm (0.1) | 0.1 μm | 0.1 μm |
| $w_l$ ($w_l^*$) | Base lead width | 0.1 μm (0.01) | 0.1 μm (0.01) | 0.05 μm | 0.05 μm |

TABLE II. List of symbols.

| Name | Dimensionless symbol | Definition |
|---|---|---|
| Length | $x^*$ | $\frac{x}{L}$ |
| Ion concentration | $c_i^*$ | $\frac{c_i}{c_b}$ |
| Fixed volume charge density | $\rho^*$ | $\frac{\rho}{c_b F}$ |
| Electric potential | $\phi^*$ | $\frac{\phi}{\phi_{\text{th}}}$ |
| Diffusion coefficient | $D^*$ | $\frac{D_i}{D_0}$ |
| Time | $t^*$ | $\frac{D_0 t}{L^2}$ |
| Ion flux | $J_i^*$ | $\frac{J_i L}{D_0 c_b}$ |
| Ionic Current | $I_i^*$ | $\frac{I_i}{F D_0 c_b}$ |
| Gradient operator | $\nabla^*$ | $L \nabla$ |

## ACKNOWLEDGEMENTS

S. T. acknowledges support from the Japan Society for the Promotion of Science (JSPS) Overseas Research Fellowship. This work was also supported by the Kayamori Foundation of Informational Science Advancement, the Tohoku Initiative for Fostering Global Researchers for Interdisciplinary Sciences (TI-FRIS) under MEXT's Strategic Professional Development Program for Young Researchers, and JSPS KAKENHI Grant Number 24K23010.FRIS) under MEXT's Strategic Professional Development Program for Young Researchers, and JSPS KAKENHI Grant Number 24K23010.